\documentclass[12pt,a4paper]{article}
\usepackage{latexsym}
\usepackage{amssymb}
\usepackage{amsthm}
\usepackage{amsmath}
\usepackage{amsfonts}
\usepackage{graphicx}

\newtheorem{theorem}{Theorem}

\newcommand{\B}{{\Bbb B}}

\renewcommand{\proof}{\noindent {\it Proof. }}

\begin{document}
\title{On the monotone complexity of the shift operator}
\date{}
\author{Igor S. Sergeev\footnote{e-mail: isserg@gmail.com}}
\maketitle

\begin{abstract}
We show that the complexity of minimal monotone circuits
implementing a monotone version of the permutation operator on $n$
boolean vectors of length $q$ is $\Theta(qn\log n)$. In particular,
we obtain an alternative way to prove the known complexity bound
$\Theta(n\log n)$ for the monotone shift operator on $n$
boolean inputs.
\end{abstract}

{\bf Introduction.} The recent paper~\cite{net19e} shows that a
plausible hypothesis from network coding theory implies a lower
bound $\Omega(n\log n)$  for the complexity of the $n$-input
boolean shift operator when implemented by circuits over a full
basis. As a corollary, the same bound holds for the multiplication
of $n$-bit numbers. (Definitions of boolean circuits and
complexity see e.g. in~\cite{weg87e}.) Curiously, nearly at the
same time an upper bound $O(n\log n)$ for multiplication has been
proved in~\cite{hv19e}. Actually, for the shift operator, the
bound $O(n\log n)$ is trivial.

The shift can be implemented by monotone circuits. Lamagna~\cite{la75e,la79e}
and independently Pippenger and Valiant~\cite{pv76e} proved that its complexity is bounded by $\Omega(n\log
n)$ with respect to the circuits over the basis $\{\vee,\,\wedge\}$.
Essentially the same bound was established by
Chashkin~\cite{ch06e} for the close problem of 
implementation of the real-valued shift operator by circuits over
the basis of 2-multiplexors and binary boolean functions. We show that the argument from~\cite{ch06e}
works for the boolean setting as well thus obtaining yet another proof of the known result. On
the other hand, an upper bound $O(n\log n)$ is easy to obtain when
a suitable encoding of the shift value is chosen.

A version of the shift operator may be seen as a partially defined
order-$n$ boolean convolution operator. It is known that the
complexity of the convolution is $n^{2-o(1)}$~\cite{gs11e}, while
the complexity of the corresponding shift operator is
$\Theta(n\log n)$.

A more general form of shift is permutation. By analogy, one can
introduce a monotone permutation operator. If a special encoding
on the set of permutations is chosen, then the permutation
operator on $n$ boolean inputs can be implemented with complexity
$O(n\log n)$ employing the optimal sorting network
from~\cite{aks83e}. If size-$n$ boolean vectors are given as
inputs, then there exists a version of the permutation operator,
which is the restriction of the $n\times n$ boolean matrix
multiplication operator. The boolean matrix multiplication
complexity is known to be $\Theta(n^3)$~\cite{me74e,pa75e}. It can be
compared with the complexity $\Theta(n^2\log n)$ of the
corresponding permutation operator. (The lower bound follows from
the bound on the complexity of the shift operator.)

{\bf Preliminaries.} Further, $L(F)$ denotes the complexity of
implementing the operator $F$ by circuits over the basis
$\{\vee,\,\wedge\}$.

Let $\B=\{0,\,1\}$ and $A = \{ \alpha_0, \ldots, \alpha_{n-1} \}
\subset \B^m$ be an antichain of cardinality $n$. By
$X=(x_0,\ldots,x_{n-1})$, $x_i = (x_{i,0}, \ldots, x_{i,q-1})^T$,
denote the $(q,n)$-matrix of boolean variables. Let
$Y=(y_0,\ldots,y_{m-1})$ denote the vector of boolean variables
encoding elements of the antichain $A$. By $v \gg k$ we denote the
vector obtained from $v$ via a cyclic shift by $k$ positions to
the right.

Monotone cyclic shift $(nq+m,nq)$-operator
$S_{q,A}(X,Y)=(s_0,\ldots,s_{n-1})$ is a partially defined
operator taking values $X \gg k$ for $Y=\alpha_k$, where
$k=0,\ldots,n-1$.

Consider a few examples of encoding shift values. The vector
$(v,\overline{v})$, where $\overline{\,\cdot\,}$ is the
componentwise negation, we call {\it doubling} of the vector $v$.
Typically, the shift value $k$ is encoded by its binary
representation $[k]_2$. For the monotone version, one can use
doubling of $[k]_2$. In this case, $m=2(\lfloor \log_2 n
\rfloor+1)$. The described encoding corresponds to the antichain
$A_0=\left\{ \left. \left( [k]_2,\,\overline{[k]_2} \right)
\right| 0 \le k < n \right\}$.

Another natural choice for $A$ is the set $A_1$ of all weight-1
vectors in $\B^n$. In this case, $m=n$. Let $q=1$. Define
$$c_i(X,Y) = \bigvee_{j + k \,=\, i \bmod n} x_j y_k.$$ The operator
$C(X,Y)=(c_0,\ldots,c_{n-1})$ is called a cyclic {\it boolean
convolution} of the vectors $X$ and $Y$.

By the definition of the shift operator, $S_{1,A_1}(X,Y)$
coincides with $C(X,Y)$ on inputs from $\B^n \times A_1$. It can
be checked that
\begin{equation*}
S_{1,A_1}(X,Y) = C(X,Y) \vee x_0\cdot \ldots \cdot x_{n-1} \cdot g
\vee r(X,Y),
\end{equation*}
where $g$ is an undefined boolean vector, and $r(X,Y)=0$ for ${|Y|
\le 1}$ (here $|v|$ denotes the weight of the vector $v$). The
complexity of convolution is known to be almost quadratic, $L(C) =
\Omega(n^2/\log^6n)$~\cite{gs11e}. Supposedly, a trivial upper
bound $L(C)=O(n^2)$ is tight. At the same time, $L(S_{1,A_1}) =
O(n\log n)$. We show below that in fact $L(S_{1,A}) = \Omega(n\log
n)$ for any $A$.

Now let $\Pi = \{ \pi_0, \ldots, \pi_{n!-1} \} \subset \B^m$ be an
antichain of cardinality $n!$. We can assign to its elements
different permutations $\pi$ on the set $\{0,\ldots,n-1\}$. Denote
$\pi(X) = \left(x_{\pi(0)},\ldots,x_{\pi(n-1)}\right)$. The
monotone permutation operator $P_{q,\Pi}(X,Y)$ is defined on
inputs $Y \in \Pi$ as $P_{q,\Pi}(X,Y) = \pi(X)$, where the
permutation $\pi$ corresponds to the value of $Y$. Since a cyclic
shift is a special case of permutation, any permutation operator
can be viewed as a shift operator defined on a larger domain.

Trivially, any permutation $\pi$ may be represented by the vector
of numbers $([\pi(0)]_2, \ldots, [\pi(n-1)]_2)$. Let $\Pi_0$
denote the corresponding coding set (it constitutes an antichain).

Otherwise, permutations may be specified as square boolean
matrices with all rows and columns having weight 1. Denote the set
of such matrices by $\Pi_1 \subset \B^{n \times n}$. The
corresponding permutation operator performs the multiplication of
the permutation matrix $Y=\{y_{j,k}\}$ by the matrix of
variables~$X$. Define
$$z_{i,k}(X,Y) = \bigvee_{j=0}^{n-1} x_{i,j} \, y_{j,k}.$$
Then $Z(X,Y) = \{ z_{i,k} \}: \B^{q \times n} \times \B^{n \times
n} \to \B^{q \times n}$ is the operator of boolean product of
matrices $X$ and $Y$. By definition, the operators $P_{q,\Pi_1}$
and $Z$ take the same values on inputs from $\B^{q \times n}
\times \Pi_1$. It is known that $L(Z) = qn(2n-1)$~\cite{pa75e}
(see also~\cite{weg87e}), which means: the naive method to
multiply boolean matrices is optimal. On the other hand,
$L(P_{q,\Pi_1}) = O(qn\log n+n^2)$ (see below). Moreover, we
manage to show that $L(P_{q,\Pi}) = \Omega(qn\log n)$ for any
$\Pi$, and this bound is achievable.

{\bf Upper complexity bounds.} For $v=(v_0,\ldots,v_{m-1}) \in
\B^m$ let $Y^v = \bigwedge_{v_i=1} y_i$ denote the monomial of
variables $y_i$ corresponding to the vector $v$. Let $L(A)$ stand
for the complexity of computation of the set of monomials $\{
Y^{\alpha} \mid \alpha \in A \}$.

\begin{theorem}
$L(S_{q,A}) \le L(A) + O(qn\log n)$.
\end{theorem}

\proof The standard circuit for the shift operator consists of
$\log_2 n$ layers of $n$ multiplexors in each. It can be built
according to the binary representation of the shift value~$k$. The
first layer shifts the input by either 0 or 1 positions, depending
on the value of the least significant bit of $k$. The second layer
shifts by 0 or 2 positions, etc.

The monotone circuit employs indicators $Y^{i,\beta} =
\bigvee_{\lfloor k/2^i\rfloor = \beta \bmod 2} Y^{\alpha_i}$ of
equality of bits of $Y$ to zeros or ones. Instead of multiplexors,
there are used similar monotone subcircuits that calculate
operators of the form $Y^{i,1} a \vee Y^{i,0} b$.

It remains to note that all boolean sums $Y^{i,\beta}$ can be
computed with complexity $O(n)$. \qed

In particular, since $L(A_0)=O(n)$ and $L(A_1)=0$, we obtain
$L(S_{1,A_0}), L(S_{1,A_1}) \in O(n\log n)$.

To derive the upper bounds on the complexity of the permutation
operator, we use a circuit $\Sigma$ sorting $n$ elements with
complexity $O(n\log n)$ provided by~\cite{aks83e}. Such a circuit
consists of comparator gates that order a pair of inputs.

\begin{theorem}

$ $

$(i)$ There exists an antichain $\Pi$ such that $L(P_{q,\Pi}) =
O(qn\log n)$.

$(ii)$ $L(P_{q,\Pi_1}) = O(qn\log n+n^2)$.
\end{theorem}

\proof A set $\Pi$ can be specified following the circuit
$\Sigma$. Assign to any permutation $\pi$ a linear order
$x_{\pi(0)} > x_{\pi(1)} > \ldots > x_{\pi(n-1)}$ on the set of
inputs of $\Sigma$ (in general, we do not consider these inputs
boolean). Let $\Sigma$ receive inputs ordered in correspondence to
a given permutation $\pi$. Assign to each comparator $e$ a boolean
parameter $y_e$ whose value is determined by the result of the
comparison. Let the doubling of the vector of parameters $y_e$, $e
\in \Sigma$, encode a permutation $\pi$.

Now, we transform the circuit $\Sigma$ to a monotone circuit for
$P_{q,\Pi}(X,Y)$, replacing any comparator $e$ receiving vector
inputs $a, b$ with a subcircuit that evaluates vectors $ay_e \vee
b\overline{y_e}$ and $a\overline{y_e} \vee by_e$.

Let us prove $(ii)$. First, recode $Y$ from $\Pi_1$ to $\Pi_0$. To
do this, one simply needs to compute positions
$y'_0,\ldots,y'_{n-1}$ of 1s in the columns of the matrix $Y$. The
position of 1 in a weight-1 column may be calculated by a trivial
circuit of linear complexity. Therefore, the complexity of the
recoding is $O(n^2)$.

Next, arrange the inputs $x_i$ in accordance to the ordering of
numbers $y'_i$ with the use of the circuit $\Sigma$. At each node
of the obtained circuit two $y'_i$ inputs are compared and,
depending on the result of the comparison, the order of the
vectors $x_i$ accompanied by the numbers $y'_i$ is determined. The
complexity of comparison is linear, so the complexity of the
subcircuit at each node is $O(q+\log n)$. \qed

{\bf Lower complexity bounds.} The proof of the following theorem
closely follows the proof of the main result in~\cite{ch06e}.

\begin{theorem}
For any choice of antichain $A$ of cardinality $n$ the following
inequality holds: $L(S_{q,A}) \ge qn\log_2 n - O(qn)$.
\end{theorem}

\proof Essentially, it suffices to consider the case $q=1$. Let
$S$ be a monotone circuit of complexity $L$ that computes
$S_{1,A}(X,Y)$.

a) First, note that for any assignment $Y=\alpha_k$ for each $i$,
the circuit $S$ contains a path connecting the input $x_i$ and the
output $s_{i+k \bmod n}$, and passing exclusively through the
gates whose outputs return the function $x_i$.

Indeed, $s_{i+k \bmod n}(X,\alpha_k) = x_i$ by definition. It
remains to check that if $x_i = f \vee g$ or $x_i = fg$, where $f$
and $g$ are monotone functions, then either $f=x_i$ or $g=x_i$.
From $x_i = f \vee g$ it follows that $f \le x_i$ and $g \le x_i$.
Assume that $f \ne x_i$ and $g \ne x_i$. It means that $f=g=0$
under the assignment $x_i=1$, $x_j=0$ for all $j \ne i$. But then
$f \vee g=0 \ne x_i$. A contradiction. The case $x_i = fg$ follows
by a dual argument.

So, moving from an output $s_{i+k \bmod n}$ towards the inputs of
the circuit, for any gate, we can select an appropriate input
computing the function $x_i$. Finally, we obtain the desired path.

b) Denote the path providing by the above argument by $p_{i,k}$.
Let $\chi(e)$ stand for the number of paths $p_{i,k}$, $0 \le i,k
< n$, passing through the gate $e$ in the circuit $S$. Note that
$\chi(e) \le n$ for all $e \in S$. Indeed, any assignment
$Y=\alpha_k$ uniquely defines the function of variables $X$
computed at the output of any gate $e$. Thus, $e$ does not belong
to two different paths $p_{i,k}$ and $p_{j,k}$. Consequently,
\begin{equation}\label{up}
\sum_{e \in S} \chi(e) \le Ln.
\end{equation}

c) Let us estimate the sum $\sum_{e \in S} \chi(e)$ in another
way. Denote by $\chi(e,j)$ the number of paths $p_{i,k}$ passing
through $e$ to the output $s_j$. By construction, $\sum_j
\chi(e,j) = \chi(e)$.

Consider the subcircuit $S_j$ obtained by combining all $n$ paths
$p_{i,k}$ leading to the output $s_j$, i.e. satisfying the
condition $i+k=j \bmod n$. By construction, $S_j$ is a connected
binary\footnote{Any vertex receives at most two incoming edges.}
directed graph with $n$ inputs and one output. We manage to bound
$\sum_{e \in S_j} \chi(e,j)$ following a simple argument
from~\cite{ls74e}\footnote{In~\cite{ls74e}, the argument was used
to bound the monotone complexity of the boolean sorting operator,
see also~\cite{weg87e}.}.

Due to the binarity property, the subcircuit $S_j$ has an input at
a distance of at least $\log_2 n$ edges from the output. In other
words, some path making up $S_j$ contains at least $\log_2 n$
gates. Exclude this path and consider a subcircuit obtained by
combining the remaining $n-1$ paths. Then, it contains a path of
length at least $\log_2 (n-1)$. We proceed this way until there is
no path remained. The argument leads to the bound
\begin{equation}\label{xej}
\sum_{e \in S_j} \chi(e,j) \ge \log_2 n! = n\log_2 n - O(n),
\end{equation} following by
\begin{equation}\label{down}
\sum_{e \in S} \chi(e) = \sum_j \sum_{e \in S_j} \chi(e,j) \ge
n^2\log_2 n - O(n^2).
\end{equation}
Putting together (\ref{up}) and (\ref{down}), we establish the
inequality $L \ge n\log_2 n - O(n)$.

d) For $q>1$, we consider separately the components of the input
and output vectors at the same positions. This results in $q$
groups of paths $p_{i,j}$. The inequality (\ref{up}) remains
valid, and the inequality (\ref{xej}) holds for any of $qn$
outputs. Thus, the required bound finally follows. \qed

Since a permutation operator is a more completely defined shift
operator, as a corollary we establish $L(P_{q,\Pi}) \ge qn\log_2 n
- O(qn)$ for any~$\Pi$.

The research is supported by RFBR grant, project
no.\,19-01-00294a.

\end{document}

\title{О монотонной сложности оператора сдвига}
\date{}
\author{И. С. Сергеев\footnote{e-mail: isserg@gmail.com}}
\maketitle

\begin{abstract}
В настоящей заметке показано, что минимальные монотонные схемы,
реализующие монотонную версию оператора перестановки $n$ булевых векторов
длины $q$, имеют сложность $\Theta(qn\log n)$. В частности, получен
альтернативный способ доказательства известной оценки $\Theta(n\log n)$ сложности
монотонного оператора сдвига булевых векторов длины $n$. % new
\end{abstract}

{\bf Введение.} В недавней работе~\cite{net19} демонстрируется,
что правдоподобная гипотеза из теории сетевого кодирования влечет
нижнюю оценку $\Omega(n\log n)$ сложности оператора сдвига булевых
векторов длины $n$ при реализации схемами в полном базисе и, как
следствие, такую же оценку для сложности умножения $n$-разрядных
чисел. (Определения булевых схем и сложности см., например,
в~\cite{weg87}.) Интересно, что практически одновременно появилась
работа~\cite{hv19}, в которой построены схемы для умножения
сложности $O(n\log n)$. Собственно для оператора сдвига оценка
сложности $O(n\log n)$ тривиальна.

Сдвиг может быть реализован монотонными схемами. Э. Ламанья~\cite{la75,la79} и  % new
независимо Н. Пиппенджер и Л. Вэльянт~\cite{pv76} доказали, что его сложность при реализации схемами над
базисом $\{\vee,\,\wedge\}$ оценивается как $\Omega(n\log n)$. Аналогичная оценка получена А. В.
Чашкиным~\cite{ch06} при решении близкой задачи реализации сдвига
действительнозначного вектора схемами над базисом из
2-мультиплексоров и бинарных булевых функций. Как будет показано
далее, практически без изменений рассуждение переносится на булев
случай и приводит к еще одной версии доказательства известного результата. % new
С другой стороны, верхнюю оценку $O(n\log n)$ несложно
получить при подходящей кодировке величины сдвига.

Доопределением одного из вариантов оператора сдвига является
оператор булевой свертки порядка $n$. Известно, что сложность
свертки имеет величину $n^{2-o(1)}$~\cite{gs11}, в то время как
сложность соответствующего оператора сдвига оказывается равной
$\Theta(n\log n)$.

Операцию сдвига обобщает операция перестановки. По аналогии, можно
определить монотонный оператор перестановки. При условии введения
специальной кодировки на множестве перестановок оператор
перестановки $n$ булевых элементов можно реализовать со сложностью
$O(n\log n)$, используя оптимальную схему сортировки
из~\cite{aks83}. Если элементы-входы представлены булевыми
векторами длины $n$, то в одном из вариантов оператор перестановки
является сужением оператора булева умножения квадратных матриц
порядка $n$. Известно, что сложность матричного умножения имеет
величину $\Theta(n^3)$~\cite{me74,pa75}, тогда как сложность
соответствующего оператора перестановки оказывается равной
$\Theta(n^2\log n)$ (нижняя оценка следует из оценки сложности
оператора сдвига).

{\bf Предварительные понятия.} Далее сложность реализации
оператора $F$ схемами над базисом $\{\vee,\,\wedge\}$ будем
обозначать через $L(F)$.

Положим $\B=\{0,\,1\}$. Пусть $A = \{ \alpha_0, \ldots,
\alpha_{n-1} \} \subset \B^m$ --- антицепь мощности $n$. Через
$X=(x_0,\ldots,x_{n-1})$, $x_i = (x_{i,0}, \ldots, x_{i,q-1})^T$,
обозначим $(q,n)$-матрицу булевых переменных. Через
$Y=(y_0,\ldots,y_{m-1})$ будем обозначать вектор булевых
переменных, кодирующий элементы антицепи $A$. Через $v \gg k$
обозначим вектор, получаемый из $v$ циклическим сдвигом на $k$
позиций вправо.

Монотонный циклический $(nq+m,nq)$-оператор сдвига
$S_{q,A}(X,Y)=(s_0,\ldots,s_{n-1})$ --- это частично определенный
оператор, принимающий значения $X \gg k$ при $Y=\alpha_k$, где
$k=0,\ldots,n-1$.

Рассмотрим примеры кодировки величин сдвига. Вектор
$(v,\overline{v})$, где $\overline{\,\cdot\,}$~--- операция
покомпонентного отрицания, назовем {\it удвоением} вектора~$v$.
Обычно величина сдвига кодируется просто своей двоичной записью. В
монотонной версии можно использовать удвоение двоичной записи. При
этом $m=2(\lfloor \log_2 n \rfloor+1)$. Если $[k]_2$ означает
двоичную запись числа $k$, то описанный способ использует антицепь
$A_0=\left\{ \left. \left( [k]_2,\,\overline{[k]_2} \right)
\right| 0 \le k < n \right\}$.

Еще один распространенный способ задания антицепи $A$ ---
множество $A_1$ всех векторов веса 1 в $\B^n$. В этом случае
$m=n$. Пусть $q=1$. Положим
$$c_i(X,Y) = \bigvee_{j + k \,=\, i \bmod n} x_j y_k.$$ Оператор
$C(X,Y)=(c_0,\ldots,c_{n-1})$ называется циклической {\it булевой
сверткой} векторов $X$ и $Y$.

По определению оператора сдвига, $S_{1,A_1}(X,Y)$ совпадает с
$C(X,Y)$ на множестве $\B^n \times A_1$. Можно проверить, что
\begin{equation*}
S_{1,A_1}(X,Y) = C(X,Y) \vee x_0\cdot \ldots \cdot x_{n-1} \cdot g
\vee r(X,Y),
\end{equation*}
где $g$ --- неопределенный булев вектор, а $r(X,Y)=0$ при условии
${|Y| \le 1}$ (через $|v|$ обозначается вес вектора $v$).
Из~\cite{gs11} известно, что сложность свертки почти квадратична,
$L(C) = \Omega(n^2/\log^6n)$, а предположительно тривиальная
верхняя оценка $L(C)=O(n^2)$ является неулучшаемой. В то же время
$L(S_{1,A_1}) = O(n\log n)$. Ниже будет показано, что на самом
деле $L(S_{1,A}) = \Omega(n\log n)$ при любом $A$.

Теперь пусть $\Pi = \{ \pi_0, \ldots, \pi_{n!-1} \} \subset \B^m$
--- антицепь мощности $n!$. Элементам антицепи можно сопоставить
различные перестановки $\pi$ на множестве $\{0,\ldots,n-1\}$.
Обозначим $\pi(X) = \left(x_{\pi(0)},\ldots,x_{\pi(n-1)}\right)$.
Монотонный оператор перестановки $P_{q,\Pi}(X,Y)$ определяется при
$Y \in \Pi$ как $P_{q,\Pi}(X,Y) = \pi(X)$, где перестановка $\pi$
соответствует значению $Y$. Поскольку циклический сдвиг ---
частный случай перестановки, любой оператор перестановки является
доопределением некоторого оператора сдвига.

Тривиально, любую перестановку $\pi$ можно закодировать вектором
двоичных кодов $([\pi(0)]_2, \ldots, [\pi(n-1)]_2)$.
Соответствующее множество кодов обозначим через $\Pi_0$ (оно
является антицепью).

Иначе, перестановки можно записывать квадратными булевыми
матрицами, в каждой строке и каждом столбце которых ровно по одной
единице. Обозначим множество таких матриц через $\Pi_1 \subset
\B^{n \times n}$. Действие соответствующего оператора перестановки
задается умножением матрицы перестановки $Y=\{y_{j,k}\}$ на
матрицу переменных $X$. Обозначим
$$z_{i,k}(X,Y) = \bigvee_{j=0}^{n-1} x_{i,j} \, y_{j,k}.$$
Оператор $Z(X,Y) = \{ z_{i,k} \}: \B^{q \times n} \times \B^{n
\times n} \to \B^{q \times n}$ выполняет булево умножение матриц
$X$ и $Y$. По построению, на множестве $\B^{q \times n} \times
\Pi_1$ значения операторов $P_{q,\Pi_1}$ и $Z$ совпадают.
Известно~\cite{pa75}, что $L(Z) = qn(2n-1)$ (см.
также~\cite{weg87}), что значит: наивный способ умножения матриц
оптимален. С другой стороны, как можно проверить, $L(P_{q,\Pi_1})
= O(qn\log n+n^2)$. Ниже будет показано, что $L(P_{q,\Pi}) =
\Omega(qn\log n)$ при любом $\Pi$, и эта оценка достижима.

{\bf Верхние оценки сложности.} Пусть $v=(v_0,\ldots,v_{m-1}) \in
\B^m$. Моном переменных $y_i$, соответствующий вектору $v$,
обозначим через $Y^v = \bigwedge_{v_i=1} y_i$. Пусть $L(A)$
означает сложность реализации системы мономов $\{ Y^{\alpha} \mid
\alpha \in A \}$.

\begin{theorem}
Справедливо соотношение $L(S_{q,A}) \le L(A) + O(qn\log n)$.
\end{theorem}

\proof Стандартная схема сдвига в полном базисе состоит из $\log_2
n$ слоев по $n$ мультиплексоров в каждом. Она строится в
соответствии с двоичной записью величины сдвига~$k$. На первом
слое, в зависимости от значения младшего разряда числа~$k$,
выполняется сдвиг на 0 или на 1 позицию. На следующем слое --- еще
на 0 или на 2 и т.д.

В монотонном случае роль двоичных разрядов играют индикаторы
$Y^{i,\beta} = \bigvee_{\lfloor k/2^i\rfloor = \beta \bmod 2}
Y^{\alpha_i}$ равенства разрядов величины сдвига нулю или единице.
Вместо мультиплексоров используются аналогичные монотонные
подсхемы, реализующие операторы вида $Y^{i,1} a \vee Y^{i,0} b$.

Остается заметить, что все булевы суммы $Y^{i,\beta}$ можно
вычислить со сложностью $O(n)$. \qed

В частности, ввиду $L(A_0)=O(n)$ и $L(A_1)=0$ получаем
$L(S_{1,A_0}), L(S_{1,A_1}) \in O(n\log n)$.

Для вывода верхних оценок сложности оператора перестановки
воспользуемся схемой $\Sigma$ сортировки $n$ элементов сложности
$O(n\log n)$ из работы~\cite{aks83}. Такая схема состоит из
элементов-компараторов, упорядочивающих пару входов.

\begin{theorem}

$ $

$(i)$ Существует антицепь $\Pi$, такая, что $L(P_{q,\Pi}) =
O(qn\log n)$.

$(ii)$ Справедливо соотношение $L(P_{q,\Pi_1}) = O(qn\log n+n^2)$.
\end{theorem}

\proof Множество $\Pi$ построим, следуя схеме $\Sigma$. Каждой
перестановке $\pi$ сопоставим линейный порядок $x_{\pi(0)} >
x_{\pi(1)} > \ldots > x_{\pi(n-1)}$ на множестве входных элементов
схемы $\Sigma$ (эти входы, вообще говоря, не считаем булевыми).
Пусть на вход схемы $\Sigma$ подан входной набор элементов,
значения которых удовлетворяют указанному порядку
(соответствующему перестановке $\pi$). Каждому компаратору $e$
сопоставим булев признак $y_e$, значение которого определяется
результатом сравнения. Перестановку $\pi$ закодируем удвоением
вектора значений признаков $y_e$, $e \in \Sigma$.

Теперь схему $\Sigma$ можно перестроить в схему для
$P_{q,\Pi}(X,Y)$, заменив каждый компаратор $e$ со
входами-векторами $a, b$ подсхемой, вычисляющей выражения $ay_e
\vee b\overline{y_e}$ и $a\overline{y_e} \vee by_e$.

Докажем $(ii)$. Перекодируем входы $Y$ из $\Pi_1$ в $\Pi_0$. Для
этого нужно просто вычислить номера $y'_0,\ldots,y'_{n-1}$ позиций
единиц в столбцах матрицы $Y$. Номер единицы в столбце вычисляется
тривиальной схемой линейной сложности, следовательно, общая
сложность перехода равна $O(n^2)$.

Далее упорядочим входы $x_i$ в соответствии с номерами $y'_i$,
используя схему $\Sigma$. В каждом узле схемы выполняется
сравнение номеров входов $y'_i$ и, в зависимости от результата
сравнения, определяется порядок векторов $x_i$ и номеров $y'_i$ на
выходе узла. Сложность сравнения линейна, поэтому обработка в
каждом узле выполняется подсхемой сложности $O(q+\log n)$. \qed

{\bf Нижние оценки сложности.} Доказательство следующей теоремы
следует доказательству основного результата~\cite{ch06}.

\begin{theorem}
При любом выборе антицепи $A$ мощности $n$ справедливо соотношение
$L(S_{q,A}) \ge qn\log_2 n - O(qn)$.
\end{theorem}

\proof Фактически достаточно рассмотреть случай $q=1$. Пусть $S$
--- монотонная схема сложности $L$, вычисляющая
$S_{1,A}(X,Y)$.

а) Заметим, что при любой подстановке вида $Y=\alpha_k$ для
каждого $i$ в схеме имеется путь, соединяющий вход $x_i$ и выход
$s_{i+k \bmod n}$, проходящий исключительно через элементы, на
выходах которых вычисляется функция $x_i$.

Действительно, $s_{i+k \bmod n}(X,\alpha_k) = x_i$ по определению.
Остается проверить, что если $x_i = f \vee g$ или $x_i = fg$, где
$f$ и $g$ --- монотонные функции, то либо $f=x_i$, либо $g=x_i$.
Из $x_i = f \vee g$ следует, что $f \le x_i$ и $g \le x_i$. Пусть
$f \ne x_i$ и $g \ne x_i$. Это значит, что $f=g=0$ при подстановке
$x_i=1$, $x_j=0$ для всех $j \ne i$. Но тогда и $f \vee g=0 \ne
x_i$. Противоречие. Случай $x_i = fg$ разбирается двойственным
образом.

Таким образом, двигаясь от выхода $s_{i+k \bmod n}$ в направлении
входов схемы, для каждого очередного элемента можно выбрать вход
со значением $x_i$. В итоге получится искомый путь.

б) Обозначим построенный в предыдущем пункте путь через $p_{i,k}$.
Пусть $\chi(e)$ обозначает число путей $p_{i,k}$, $0 \le i,k < n$,
проходящих через элемент $e$ в схеме $S$. Заметим, что $\chi(e)
\le n$ для всех $e \in S$. Действительно, подстановка $Y=\alpha_k$
однозначно задает функцию переменных $X$, реализуемую на выходе
любого элемента $e$, поэтому через $e$ не могут проходить два
разных пути $p_{i,k}$ и $p_{j,k}$. Таким образом,
\begin{equation}\label{up}
\sum_{e \in S} \chi(e) \le Ln.
\end{equation}

в) Оценим сумму $\sum_{e \in S} \chi(e)$ другим способом.
Обозначим через $\chi(e,j)$ число путей $p_{i,k}$, проходящих
через $e$ и ведущих к выходу $s_j$. По построению, $\sum_j
\chi(e,j) = \chi(e)$.

Рассмотрим подсхему $S_j$, получаемую объединением всех $n$ путей
$p_{i,k}$, ведущих к выходу $s_j$, т.е. удовлетворяющих условию
$i+k=j \bmod n$. По построению, подсхема $S_j$ имеет вид связного
бинарного (не более двух входящих ребер у каждой вершины)
ориентированного графа с $n$ входами и одним выходом. Оценим сумму
$\sum_{e \in S_j} \chi(e,j)$, используя простое рассуждение
из~\cite{ls74}. (В работе~\cite{ls74} оценивалась монотонная
сложность оператора сортировки булевого набора, см.
также~\cite{weg87}.)

Ввиду свойства бинарности, в подсхеме $S_j$ имеется вход,
удаленный от выхода на расстояние не менее $\log_2 n$ ребер. Иначе
говоря, один из путей, образующих схему, проходит как минимум
через $\log_2 n$ элементов. Исключим этот путь и рассмотрим
подсхему, полученную объединением оставшихся $n-1$ путей. В ней
найдется путь длины не менее ${\log_2(n-1)}$. Будем продолжать в
том же духе до исчерпания множества путей. Рассуждение приводит к
оценке
\begin{equation}\label{xej}
\sum_{e \in S_j} \chi(e,j) \ge \log_2 n! = n\log_2 n - O(n),
\end{equation} из которой следует
\begin{equation}\label{down}
\sum_{e \in S} \chi(e) = \sum_j \sum_{e \in S_j} \chi(e,j) \ge
n^2\log_2 n - O(n^2).
\end{equation}
Сопоставляя (\ref{up}) и (\ref{down}), получаем соотношение $L \ge
n\log_2 n - O(n)$.

г) При $q>1$ рассматриваем по отдельности компоненты
векторов-входов и векторов-выходов на одинаковых позициях. При
этом возникает $q$ комплектов путей $p_{i,j}$. Неравенство
(\ref{up}) сохраняет силу, а неравенство (\ref{xej}) выполняется
для каждого из $qn$ выходов, откуда следует окончательная оценка.
\qed

Так как оператор перестановки является доопределением оператора
сдвига, в качестве следствия получаем $L(P_{q,\Pi}) \ge qn\log_2 n
- O(qn)$ при любом $\Pi$.

Работа выполнена при поддержке РФФИ, проект 19-01-00294а.